\documentclass[fleqn,10pt]{wlscirep}
\usepackage[utf8]{inputenc}
\usepackage[T1]{fontenc}
\usepackage{subfiles}
\usepackage{multirow}
\usepackage{hyperref}

\newcommand{\comment}[1]{}
\newcommand\rebuttal[1]{\textcolor{black}{#1}}

\title{\rebuttal{Addressable Superconductor Integrated Circuit Memory from Delay Lines}}

\author[1,2,*]{Jennifer Volk}
\author[2]{Alex Wynn}
\author[3]{Timothy Sherwood}
\author[4,*]{Georgios Tzimpragos}
\affil[1]{UC Santa Barbara, Department of Electrical and Computer Engineering, Santa Barbara, CA, 93106, USA}
\affil[2]{Massachusetts Institute of Technology, Lincoln Laboratory, Lexington, MA, 02420, USA}

\affil[3]{UC Santa Barbara, Department of Computer Science, Santa Barbara, CA, 93106, USA}
\affil[4]{University of Michigan, Department of Electrical Engineering and Computer Science, Ann Arbor, MI, 48109, USA}
\affil[*]{jevolk@ucsb.edu, gtzimpra@umich.edu}

\begin{abstract}
Recent advances in logic schemes and fabrication processes have renewed interest in using superconductor electronics for energy-efficient computing and quantum control processors. However, scalable superconducting memory still poses a challenge. To address this issue, we present an alternative to approaches that solely emphasize storage cell miniaturization by exploiting the minimal attenuation and dispersion properties of superconducting passive transmission lines to develop a delay-line memory system. This fully superconducting design operates at speeds between 20 GHz and 100 GHz, with $\pm$24\% and $\pm$13\% bias margins, respectively, and demonstrates data densities in the 10s of Mbit/cm$^2$ with the MIT Lincoln Laboratory SC2 fabrication process. Additionally, the circulating nature of this design allows for minimal control circuitry, eliminates the need for data splitting and merging, and enables inexpensive implementations of sequential access and content-addressable memories. Further advances in fabrication processes suggest data densities of 100s of Mbit/cm$^2$ and beyond.

\end{abstract}
\begin{document}

\flushbottom
\maketitle

\thispagestyle{empty}

\section*{Introduction}
Superconductor electronics (SCEs) feature almost zero static power dissipation, speed-of-light energy-efficient interconnects, and clock rates in the 100s of GHz~\cite{7368011}. In addition to these characteristics, SCEs can serve as facilitators for integrated classical-quantum computers due to their cryogenic nature~\cite{PhysRevApplied.12.014044, 9139002}. Despite advances in fabrication~\cite{8314189}, tools~\cite{8607085, christensen2022pylse}, and logic schemes~\cite{ctl, xsfq}, 
however, the lack of a reliable high-speed and high-density superconducting memory continues to impede the development of practical SCE systems~\cite{6449287}. In this paper, we introduce a scalable superconducting delay line memory that takes advantage of the technology's fast switching, zero resistance, and high-kinetic inductance properties.

Previous research has shown that directly applying single flux quantum (SFQ) principles to memory results in designs with low access latency but insufficient density~\cite{783834}. While arrays of vortex transition (VT) cells have demonstrated the ability to store up to 1 Mbit of data per square centimeter~\cite{8667388}, they face significant limitations for further advancement due to their reliance on superconducting transformers. On the other hand, hybrid architectures that combine SFQ and complementary metal–oxide semiconductor (CMOS) technologies provide better scalability, albeit with long access latencies~\cite{6363561}. CMOS units may scale more effectively than their superconducting counterparts but are slower and usually reside outside of the 4.2 kelvin cryocooler due to their power and thermal footprints. In the search for a viable superconducting memory solution, a considerable effort has also been invested in superconducting memory cells built from novel superconducting-ferromagnetic stack-ups~\cite{Feofanov_2010, 6377273, Baek_2014, Gingrich_2016, osti_10153980}. While promising in many aspects, such designs suffer from complex device structures and thus have their own practical limitations. Lastly, an approach that attempts to find a compromise between the hard-to-scale VT cells and the hard-to-fabricate superconducting-ferromagnetic hybrids is that of superconducting nanowire memory cells~\cite{Murphy_2017, Zhao_2018}. Recent implementation results indicate a bit cell area of 26.5~$\mu$m$^2$~\cite{Butters_2021}, which is the most compact experimentally-verified superconducting storage element to date. However, despite their advantages, superconducting nanowire memory cells are addressed by hTrons, which bring about relatively slow access times, considerably high power consumption, and high error rates for multi-cell arrays.

\begin{figure}[ht]
\centering
\includegraphics[width=0.40\linewidth]{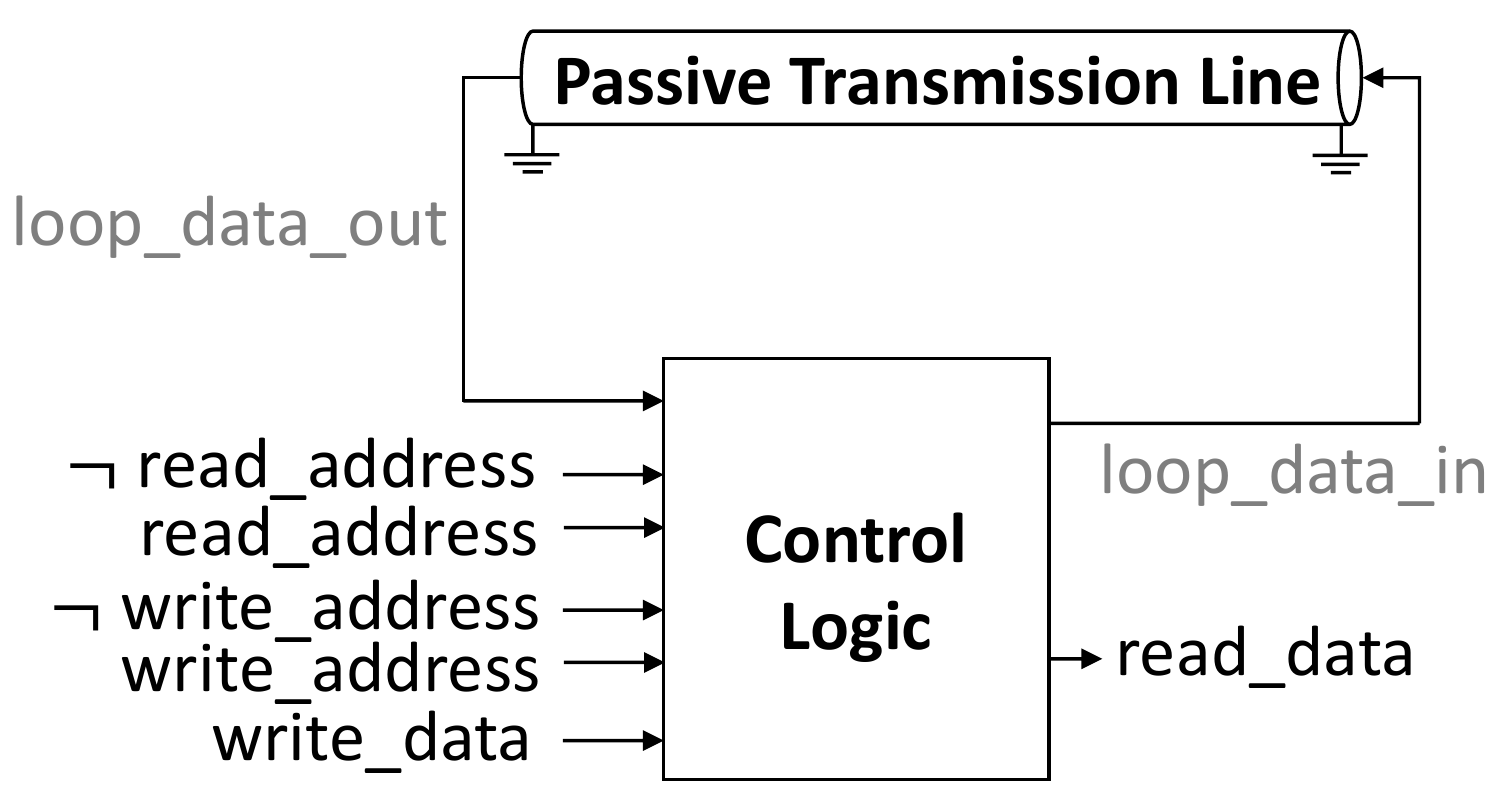}
\caption{\rebuttal{High-level view of the proposed memory design. The storage loop is implemented with a passive transmission line (PTL). SFQ pulses introduced at one end of the PTL ($loop\_data\_in$) travel along it at a controlled speed for a given time and get picked up at the output end of the line ($loop\_data\_out$).
The memory control logic decides whether received pulses will be read out and whether they will be forwarded again to the input of the loop to repeat the cycle.}}
\label{fig:arch1}
\end{figure}

In this paper, we forgo traditional array structures and propose a novel superconducting delay-line memory based on Passive Transmission Lines (PTLs). PTLs are, in essence, superconducting wires that can transmit single flux quanta with zero resistance, making them ideal for signal routing in large-scale SFQ designs~\cite{4277484}. If used to form loops, however, they can also be thought of as storage mediums. Figure~\ref{fig:arch1} illustrates one such loop, which transmits SFQ pulses introduced at its input and delivers them after a specific time to its output. The pulses are then picked up and fed back to the loop input, where they are circulated again. Reading from and writing to the memory is achieved in a time-serial manner. This enables the time-sharing of control circuitry, eliminates the need for data splitting and merging, and, in doing so, circumvents the obstacles presented by previous approaches.

The idea of using delay lines in memory implementations has roots that go back several decades. One of the earliest examples was the acoustic delay line memory, invented by Eckert and Mauchly in 1947~\cite{eckert1953memory}, and published by Auerbach, Eckert, et al. in 1949~\cite{1698100}. In the mid-1950s, IBM produced the 650 calculator, which relied on a magnetic delay line memory known as drum memory~\cite{10.1145/320764.320768, 6442018}. 
Approximately a decade later, Oregon State University designed NEBULA, a medium-speed serial digital computer that utilized a content-addressable memory (CAM) made from 35 glass delay lines~\cite{1671296}. IBM also experimented with a similar concept in the 1970s while constructing non-volatile bubble memory. Bubble memory stores data in small magnetized ``bubble'' areas, and read and write operations are performed by repositioning these bubbles~\cite{1067599}. More recently, racetrack memories have been developed, which promise to deliver high performance at a low cost. In this case, data is stored along a series of magnetic domain walls in non-superconducting magnetic nanowires. Reading and writing are accomplished by passing current through the nanowires, which forces domain walls to advance~\cite{Parkin190, Parkin2015}.

When it comes to superconductor electronics, the concept of realizing memory through data circulation is nearly unexplored. Hattori et al.~\cite{783862} made an early attempt to utilize a YBa$_2$Cu$_3$O$_{7-\delta}$ (YBCO) coplanar delay line in the late 1990s for constructing high-speed cell buffer storage for asynchronous transfer mode switching systems. Another effort was made in 2016 by Ishida et al.~\cite{7799755}, who proposed an SFQ cache architecture that relies on circular shift registers built from synchronous destructive read-out (DRO) cells. Although these approaches provide evidence of the feasibility and speed of superconducting delay line memories, their underlying designs are limited in capacity or addressing capability. For example, in the study conducted by Hattori et al.~\cite{783862}, a GaAs 2$\times$2 crossbar switch allowed for an interface to be established with the YBCO coplanar delay line. But this crossbar \rebuttal{lacks support for addressing, non-destructive readouts, or speeds exceeding 10~GHz.}
In addition, the estimated signal travel speed in the YBCO line is approximately $0.4c$, where $c$ is the speed of light. This implies that the minimum spacing between two subsequent data signals in the YBCO line would be about 12~mm. \rebuttal{Hence, both the controller design and the line material impose significant constraints on the memory density.} Regarding the DRO-based cache architecture~\cite{7799755}, synchronous cells are used to form a circular shift register, which hampers energy and area efficiency. Furthermore, shifting is controlled by a sequence of clock pulses whose length is determined by the provided address and the register’s current position, thereby resulting in additional overhead. 

The proposed PTL-based memory design, in contrast, employs primarily passive components; is \rebuttal{fully superconducting}; encompasses all typical memory functionalities\rebuttal{, such as addressing, data overwrite, and non-destructive readout}; and achieves interface speeds of up to 100 GHz in simulation with satisfactory bias margins. \rebuttal{The high controller (pulse injection) speed, coupled with the SFQ slow-down caused by high-kinetic inductance materials, boosts memory density by reducing the minimum spacing between subsequent pulses in PTLs. For example, the minimum spacing in NbN nanowires is about 570 times shorter than that of YBCO lines~\cite{783862}.}
To validate these hypotheses and quantify the projected gains, we conduct detailed analog simulations and formulate models that establish the relationship between memory density and various factors such as the operating frequency of the interface circuitry, pulse travel speed in the PTL, and line dimensions. For the latter two, we investigate PTL designs with different topologies, material compositions, and fabrication processes.

\section*{Results}
\subsection*{Architecture Description and Functional Evaluation}
The block diagram of the proposed memory is shown in Figure~\ref{fig:arch1}. The design consists of a PTL-based delay line and a control logic block. The delay line serves as the circulating loop storage and delays any data that arrives at its input (\emph{loop\_data\_in}). The delay introduced by the loop depends on the line's length and the pulse travel speed in the line. At the end of each round trip, the data at the output of the line (\emph{loop\_data\_out}) enters the controller, which serves as a memory interface. The controller is responsible for deciding whether signals from the feedback path (\emph{loop\_data\_out}) or the input (\emph{write\_data}) will be forwarded to the delay line (\emph{loop\_data\_in}) for another round and whether it will be copied and forwarded to the readout port (\emph{read\_data}).

\begin{figure}[ht]
\centering
\includegraphics[width=0.52\linewidth]{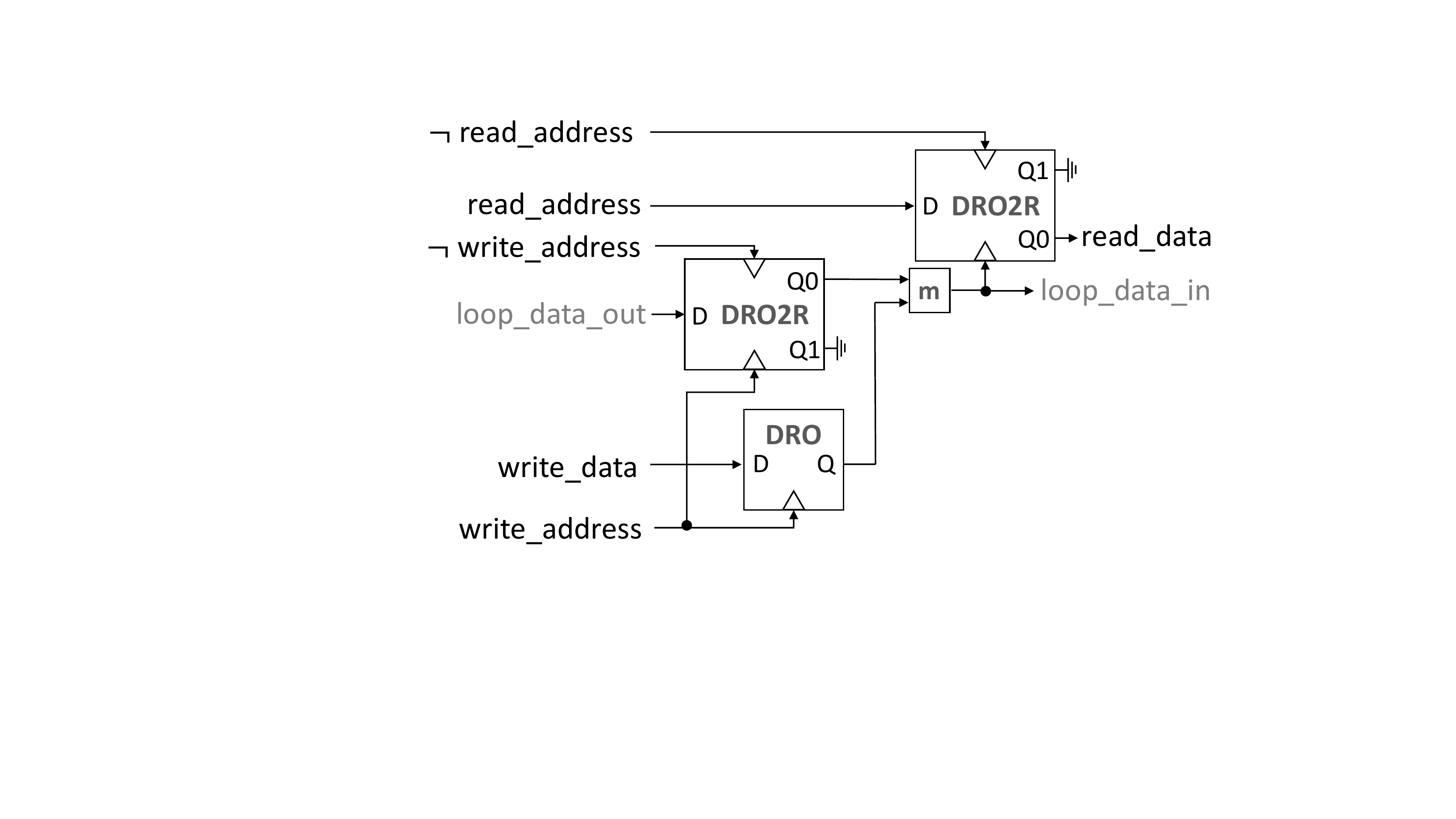}
\caption{Block diagram of a controller enabling sequential-access addressing. A destructive readout (DRO) cell stores a $write\_data$ pulse until $write\_address$ arrives. A DRO cell with two readout ports (DRO2R) is used to re-synchronize data on $loop\_data\_out$ with the arrival of a pulse on the $\neg{write\_address}$ line; or, conversely, $loop\_data\_out$ is filtered out upon the arrival of $write\_address$. For reading, another DRO2R  cell is used. In this case, a pulse on the $read\_address$ line loads this DRO2R cell, and a subsequent $loop\_data\_in$ or $\neg{read\_address}$ pulse triggers it. If  $loop\_data\_in$ arrives first, the stored pulse is forwarded to the $read\_data$ line; otherwise, it is filtered out. A Merger cell, denoted by $m$, joins two lines: the first from the output of the DRO for a write operation, and the second from the DRO2R for memory loop re-circulation. A fan-out of two is required after the Merger; cell $I_C$ ranking is adopted to maintain signal fidelity around this point~\cite{ranking}. \rebuttal{The final design uses serial resistors at the boundaries of the PTL to eliminate stray DC currents caused by trapped fluxons and to provide damping for possible reflections due to impedance mismatches.}}
\label{fig:arch2}
\end{figure}

Figure~\ref{fig:arch2} illustrates the schematic of the controller. Temporally-encoded SFQ signals, generated by comparing the value of an address counter with a target address, are used for addressing. The Merger cell, denoted by the letter \emph{m}, stitches together and forwards all signals that appear on its two input lines to its single output line. When no pulse arrives on the \emph{write\_address} signal line within the designated interval, a pulse appears on its complementary signal line, \emph{$\neg{write\_address}$}. A pulse on the \emph{loop\_data\_out} line then flows from the DRO2R (DRO with two outputs) on the left into the delay line input (\emph{loop\_data\_in}) on the right, without waiting for other signals. Inversely, when a pulse arrives on the \emph{write\_address} line, a pulse on the \emph{loop\_data\_out} line is ignored, the content of the DRO2R is cleared, and \emph{write\_data} is forwarded to the delay line input and readout circuitry. The use of differential signaling for write addressing enables the correction of potential data timing distortions incurred in the control circuitry and the storage loop. The readout circuitry on the right comprises a DRO2R cell. For readout to occur, a pulse is loaded through the \emph{read\_address} line. As with the first DRO2R, there are two cases: either a pulse arrives through the \emph{loop\_data\_in} line and pushes the stored value to the Q0 output port (\emph{read\_data}), or a pulse on the complementary \emph{$\neg{read\_address}$} line clears the cell, flushing the stored value.

\begin{figure}[ht]
\centering
\includegraphics[width=0.95\linewidth]{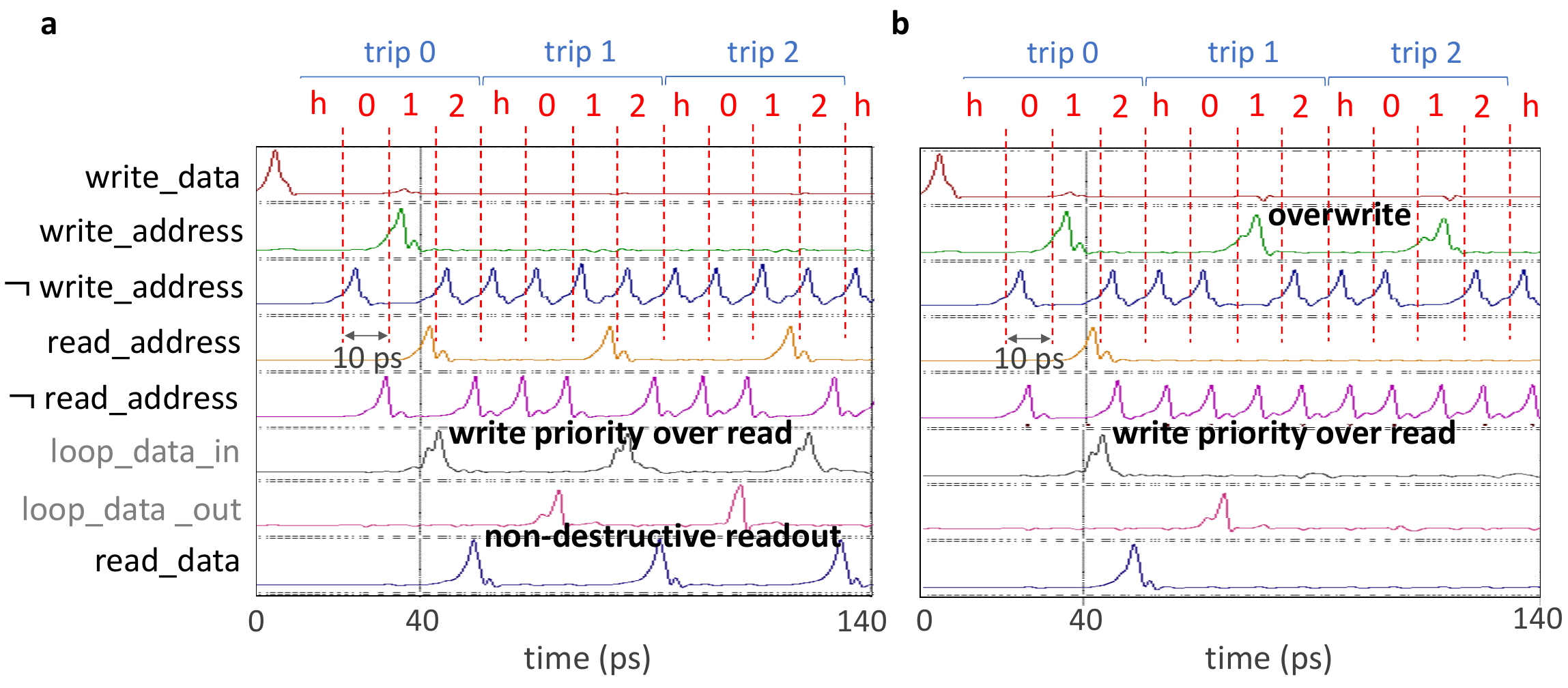}
\caption{WRSPICE~\cite{Wrspice} simulation results of the presented superconducting delay-line memory using MIT Lincoln Laboratory's 100~$\mu A/\mu m^2$ SFQ5ee process parameters. The operating frequency is set to 100~GHz, and the number of supported addresses is set to three. A header interval, $h$, is needed at the beginning of each memory rotation, or trip, to load data for writing. \rebuttal{Panel a}: The loop starts empty and a pulse on the \emph{write\_data} signal line is provided in the header interval of trip 0. During the third interval of the same trip, a pulse is observed on both the \emph{write\_address} and \emph{read\_address} lines, indicating a write operation to and read operation from address 1. After the \emph{write\_address} pulse arrival, a pulse appears on the \emph{loop\_data\_in} line, demonstrating a successful memory write operation. After the \emph{read\_address} pulse arrives, a pulse is observed on the \emph{read\_data} line, indicating a successful memory read operation. During the following two trips, the pulse stored in address 1 is successfully retrieved again, which provides evidence of the memory's non-destructive readout capability. \rebuttal{Panel b}: The first rotation, trip 0, is identical to that of \rebuttal{Panel a}. However, in trip 1, the pulse stored in address 1 is successfully overwritten, as indicated by the absence of pulses in the \emph{loop\_data\_in} and \emph{loop\_data\_out} lines after the overwrite.}

\label{fig:func_eval}
\end{figure}

Simulation results for this design are provided in Figure~\ref{fig:func_eval}. In both cases, the controller operates at 100~GHz, and three full rotations, or trips, are shown. Each trip consists of four intervals, with three of them corresponding to the number of supported memory addresses and one serving as the header, denoted by \emph{h}. More specifically, in Figure~\ref{fig:func_eval}a, a pulse is provided on the \emph{write\_data} line during the header interval of the first rotation, trip 0. In the third interval of the same trip, a pulse arrives on each of the \emph{write\_address} and \emph{read\_address} lines, denoting a write to and read from address 1. Upon the arrival of the \emph{write\_address} pulse, a pulse appears on the \emph{loop\_data\_in} line, which demonstrates a successful memory write operation. The subsequent appearance of a pulse on the \emph{read\_data} line after the arrival of the pulse on the \emph{read\_address} line indicates that write operations have higher priority than read. To illustrate the non-destructive nature of readout, in trip 1 of Figure~\ref{fig:func_eval}a, a pulse is asserted on the \emph{read\_address} line again, but this time, it is not paired with a pulse on the \emph{write\_address} line. It is again followed by the appearance of a pulse on the \emph{read\_data} line, which evidences the desired behavior. To demonstrate data overwrite, the same pulse ordering as before is used to set up the memory in Figure~\ref{fig:func_eval}b. However, in this case, a pulse is asserted on the \emph{write\_address} line during trip 1 without being accompanied by a second \emph{write\_data} pulse. As anticipated, no pulse appears on the \emph{loop\_data\_in} line after this operation, indicating that the first pulse was successfully overwritten.

Note that the presented memory system allows one to search and operate on all of the memory contents while waiting for the entire circulation time to pass, thereby eliminating the need to broadcast to or continuously poll individual cells. The design's rotating nature not only circumvents classic fan-in and fan-out limitations of superconductor electronics but also supports the addition of multiple write and read ports and the inexpensive implementation of content-addressable memories~\cite{1671296}.

\subsection*{Circuit Descriptions and Performance Evaluation}
\begin{figure}[ht]
\centering
\includegraphics[width=\linewidth]{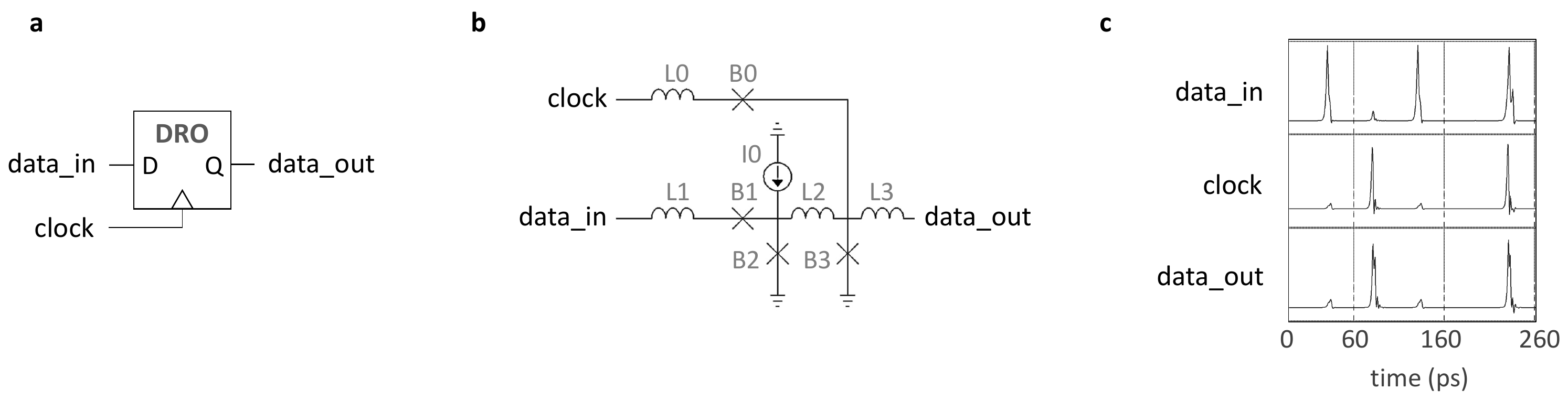}
\caption{\rebuttal{Symbol (Panel a), schematic (Panel b), and simulation results (Panel c) of a destructive readout (DRO) cell. An incoming $data\_in$ SFQ pulse is stored in the superconducting quantum interference device (SQUID) formed by B2-L2-B3 until a $clock$ pulse arrives. The arrival of a $clock$ pulse switches B3 and releases an SFQ pulse on $data\_out$.}}
\label{fig:schem1}
\end{figure}

\begin{figure}[ht]
\centering
\includegraphics[width=\linewidth]{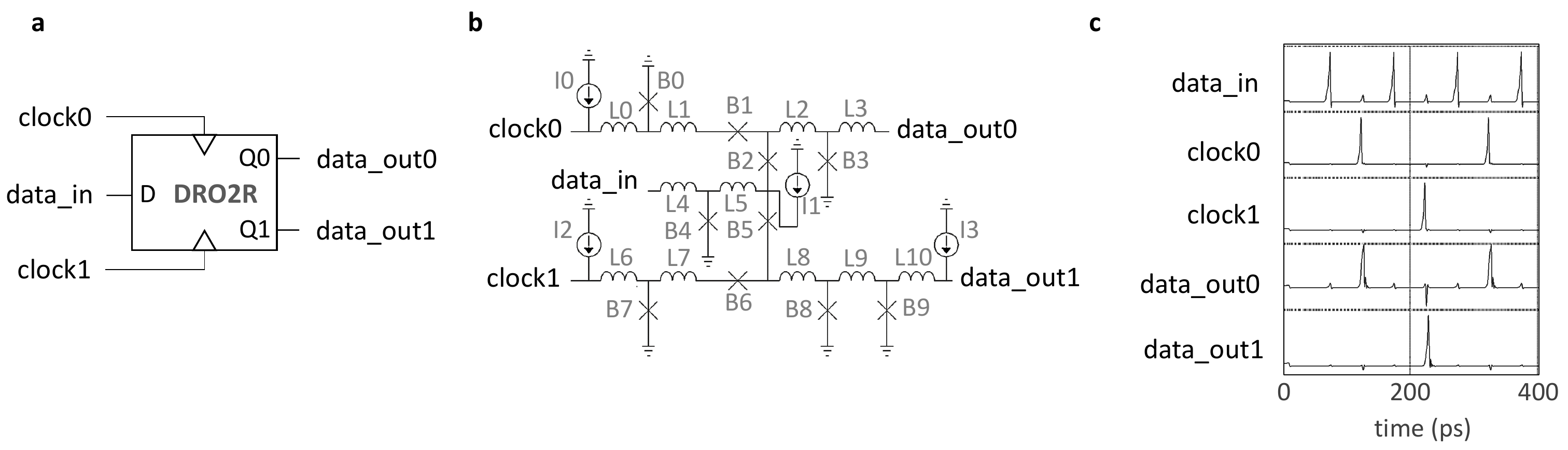}
\caption{\rebuttal{Symbol (Panel a), schematic (Panel b), and simulation results (Panel c) of a DRO cell with two readout ports (DRO2R). The DRO2R cell performs largely the same operation as the DRO---but in this case, the storage element is shared between two parallel loops: B4-L5-B2-L2-B3 and B4-L5-B5-L8-B8~\cite{dro2rdesign}. A pulse appearing on either clock input will clear the stored SFQ and push it to the respective output line.}}
\label{fig:schem2}
\end{figure}

\begin{figure}[ht]
\centering
\includegraphics[width=\linewidth]{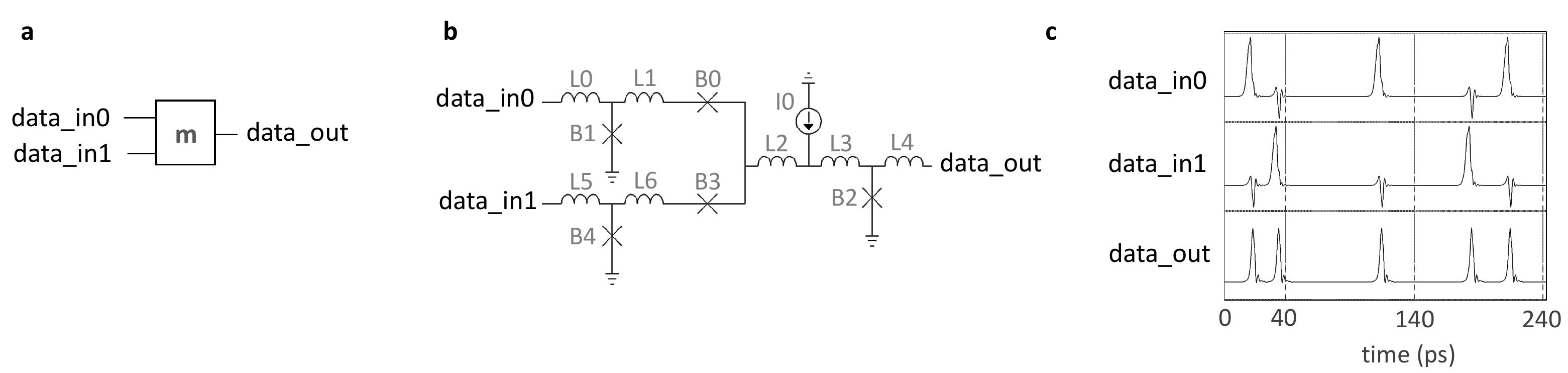}
\caption{\rebuttal{Symbol (Panel a), schematic (Panel b), and simulation results (Panel c) of a Merger cell. As its name implies, this design passes incoming SFQ pulses from either of its two input ports to its output. To prevent backward propagation of an input SFQ from the opposite line, two blocking Josephson junctions, B0 and B3, are  used.}}
\label{fig:schem3}
\end{figure}

To evaluate the performance and feasibility of the proposed memory design, we first provide schematics and simulation results for the memory controller's main components; next, we analyze their latency; lastly, we perform a voltage bias margin analysis for the entire system, including all loading effects due to the control logic, PTL, and accompanying driver and receiver circuitry. The controller, shown in Figure~\ref{fig:arch2}, consists of a DRO cell, two DRO2R cells, and a Merger cell (m). Figures~\ref{fig:schem1}-\ref{fig:schem3} provide schematics for each cell as well as corresponding simulation waveforms to demonstrate cell function.

As is the case for any system, the electrical and timing properties of these cells affect both the performance and functionality of the proposed memory. In particular, electrical issues, typically brought on by susceptibility to parametric variation, can lead to fatally under- or over-biased Josephson Junctions (JJs), which in turn can lead to circuit dysfunction, or delayed or early switching times. To avoid erroneous behavior and ensure correct system timing, the effects of under- and over-biasing are first examined at the cell level. Performing this bias analysis for cells in isolation, however, is not sufficient because it excludes the loading effects that are present in a system setting.

\begin{figure}[ht]
\centering
\includegraphics[width=0.85\linewidth]{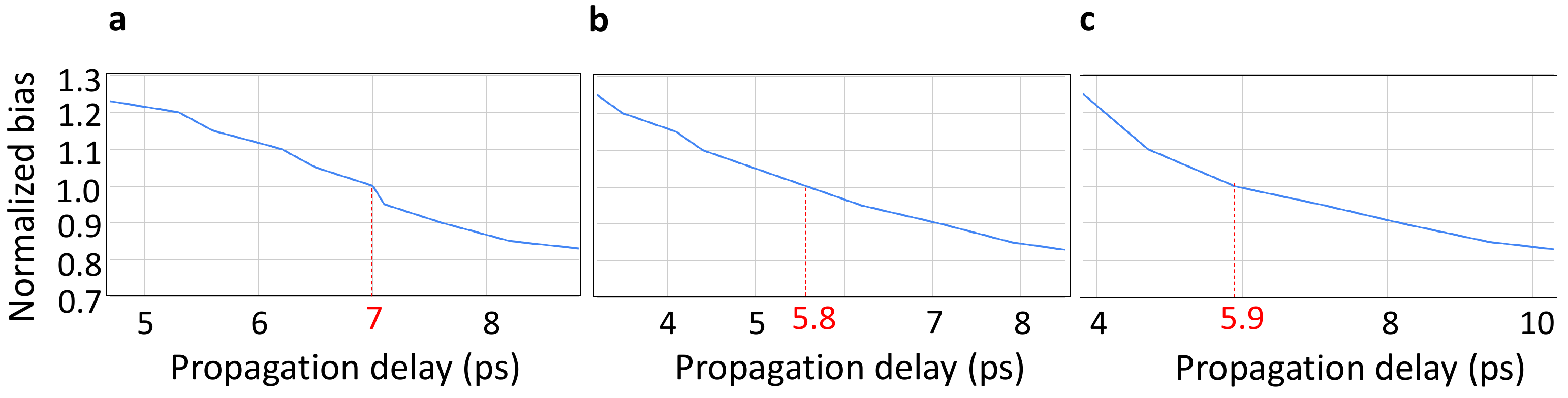}
\caption{Bias versus propagation delay for DRO \rebuttal{(Panel a)}, DRO2R \rebuttal{(Panel b)}, and Merger \rebuttal{(Panel c)} cells. 
Red markings indicate nominal values. To achieve symmetric bias margins, the nominal delay of each cell is centered between its upper and lower time bounds.}
\label{fig:bias_vs_delay}
\end{figure}

To account for loading, iterative measurements and component tuning are carried out in an in-situ approach to achieve the desired timing.
Accordingly, each cell is fully loaded by the remaining components in the memory controller.
The results of this process are shown in Figure~\ref{fig:bias_vs_delay}. Nominal cell delays are indicated in red. DRO and DRO2R delays are measured as clock-to-Q delays, while the Merger delay is measured as the propagation delay from either of the inputs to the output. Delays in each cell increase as bias decreases, and decrease as bias increases. To make bias margins symmetric, a set of component parameters is chosen that centers the nominal delay of each cell between the upper and lower time bounds of the controller.

\begin{figure}[ht]
\centering
\includegraphics[width=0.3\linewidth]{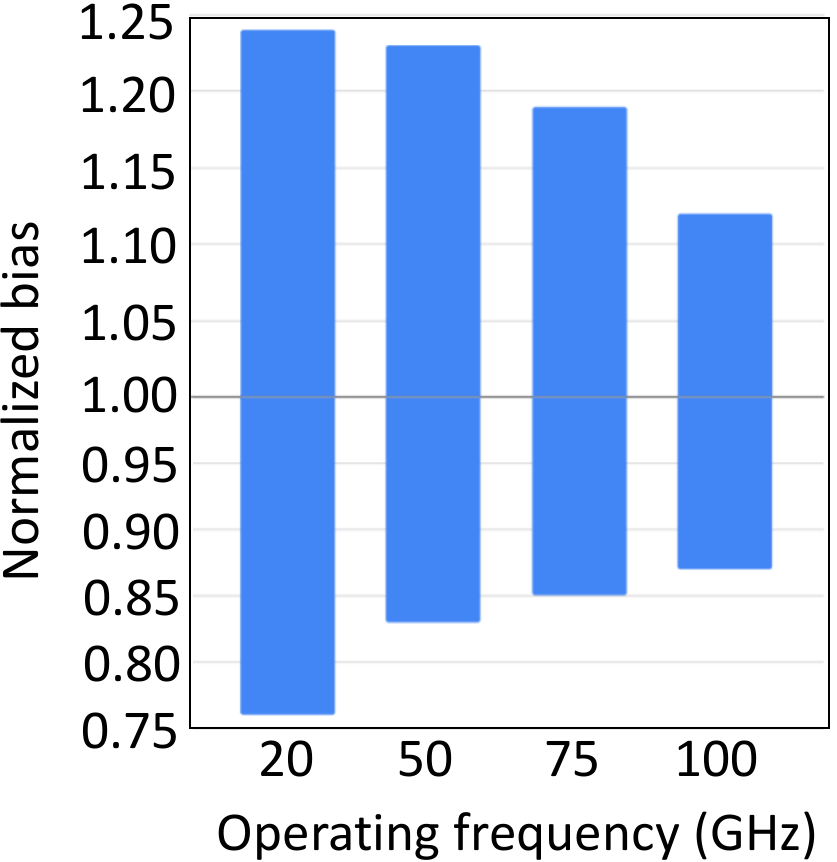}
\caption{Bias margins of the complete memory design for a variety of operating frequencies.}
\label{fig:system_bias}
\end{figure}

Using interval analysis and the above delays, the maximum operating frequency of the controller is estimated, and the cell tuning and bias margin measurements are repeated. Another round of bias margin measurements is conducted, wherein the bias of the entire design is varied instead of individual cells.
Figure~\ref{fig:system_bias} illustrates our results for frequencies ranging from 20~GHz to 100~GHz. We notice that electrical issues---caused by, for example, Josephson junctions that are subject to the above biasing concerns and that switch too frequently or not at all---drive limitations in bias margin width at lower frequencies, while timing issues are the limiter at higher frequencies. This happens because timing constraints get tighter as the address timing interval is reduced. For example, at 100~GHz, the address timing interval is just 10~ps, which leaves little room for the same variations in propagation delay that we observed in Figure~\ref{fig:bias_vs_delay}. Our SPICE simulations show bias margins ranging from $\pm$24\% (at 20~GHz) to $\pm$13\% (at 100~GHz), which surpass the widely-accepted $\pm$10\% threshold~\cite{1504851}.

\subsection*{Data Density Estimation}

The physical storage density---that is, bits per area---of the proposed memory depends on 1) the PTL's linewidth and spacing requirements, set by the fabrication process; 2) the travel speed of SFQ pulses in the PTL, set by the material of choice and the line topology; 3) the relative timing between two adjacent SFQ pulses, set by the controller's operating frequency; and 4) the number of PTL memory routing layers. We estimate the density of the proposed PTL-based superconducting delay line memory by choosing various settings for each of these free variables and summarize our results in Table~\ref{tab:tab1}.

A typical Nb stripline of 250~nm linewidth with a minimum spacing of 250~nm~\cite{8314189} propagates SFQ pulses at a speed of $0.3c$. This leads to data densities of up to 0.9~Mbit/cm$^2$ at 100~GHz, if four metal routing layers are used. Reducing the Nb stripline linewidth and minimum spacing from 250~nm to 120~nm is a possible but more aggressive design choice~\cite{Tolpygo_2021} and results in densities of up to 1.9~Mbit/cm$^2$.
By switching device material and topology to an MoN kinetic inductor microstrip with the same dimensions, available on just one layer within MIT Lincoln Laboratory's SFQ5ee~\cite{7386652} and SC2~\cite{Tolpygo_2021} processes, the travel speed of pulses in the line falls by about 6$\times$. This slowdown yields densities of up to 1.4 and 4.0~Mbit/cm$^2$ for 250~nm and 120~nm linewidths, respectively, at 100~GHz. By default, the SFQ5ee and SC2 fabrication processes allow for one MoN high-kinetic inductance layer, four Nb signal-routing layers, and three ground planes, based on the typical allocation. An increase in the number of the MoN high-kinetic inductance layers from one to four transforms the line topology into that of a stripline and increases the data density to 3.2~Mbit/cm$^2$ at 20~GHz and 19~Mbit/cm$^2$ at 100~GHz for a 120~nm linewidth and 120~nm spacing.

At this point, it is evident that the use of materials with increasingly high kinetic inductance is conducive to higher densities. To this end, we explore the potential of NbTiN striplines that exhibit approximately an order of magnitude higher inductance than their MoN counterparts and propagate SFQ pulses at a speed of $0.011c$~\cite{PhysRevLett.122.010504}. Our results indicate that for a NbTiN stripline with 100~nm width, 120~nm spacing, four metal routing layers, and controller frequencies between 20 and 100~GHz, the estimated data densities range from 10.7 to 53.3~Mbit/cm$^2$.  
A more forward-looking approach comes from the use of NbN kinetic inductor nanowires. In the case of an experimentally-tested NbN nanowire with 40~nm linewidth, the inductance scales to 2,050~pH/$\mu$m
~\cite{PhysRevApplied.11.044014}. A roughly proportional drop in capacitance keeps the pulse travel speed the same, $0.011c$, but the reduced linewidth pushes the maximum data density to 75.4~Mbit/cm$^2$ at 100~GHz. A reduction in linewidth to 15~nm causes the inductance to increase to 5,467~pH/$\mu$m~\cite{PhysRevApplied.11.044014} based on a factor of 82~pH/$\square$, which drops the travel speed to $0.007c$ and increases data density to 140.3~Mbit/cm$^2$ at 100~GHz. This is equivalent to a physical pulse spacing of 21~$\mu$m, about 600$\times$ shorter than that in the YBCO line~\cite{783862}. Moreover, assuming that CMOS layer stacking techniques---such as those used to create 100-layer stacks in V-NANDs---could be applied to superconductor technology, this NbN nanowire technology could provide a memory density of 3,507~Mbit/cm$^2$ at 100~GHz operating speed.

\section*{Discussion}
The microarchitectural features of our most prominent computer systems have historically co-evolved with the technology and devices that embody them. For instance, in the case of CMOS, the combination of cheap transistors and power-hungry interconnects~\cite{10.1145/966747.966750} has motivated the development of designs that trade off resource redundancy for reduced data movement. By contrast, data movement in superconductor electronics is relatively cheaper than switching due to the nearly lossless nature of superconducting interconnects. Therefore, the critical question---which the presented research addresses---is whether, in the case of superconductor electronics, we can capitalize on the unique properties of superconducting interconnects to improve hardware efficiency, simplicity, and density.

Previous attempts to develop cryogenic memory technology, which functions at temperatures of 4 kelvin or below, have predominantly centered on creating individual storage cells that can subsequently be arranged in a standard grid configuration~\cite{alam2023cryogenic}. While such configurations have been effective in CMOS, they cannot fully leverage the advantage that superconductor electronics provide in terms of low-cost signal transmission. Additionally, they often disregard the forthcoming obstacles that arise from the limitations of inductor scaling and fan-in/out~\cite{8667388}, as well as the significant power consumption and long access times~\cite{6363561}, high device complexity~\cite{Feofanov_2010, 6377273, Baek_2014, Gingrich_2016, osti_10153980}, and increased bit-error rate~\cite{Murphy_2017, Zhao_2018, Butters_2021} that come with the suggested structures.

The presented approach departs from this tradition and exploits superconducting passive transmission lines for the construction of a delay-line memory system. From a device perspective, delay lines are easy to construct and offer high signal fidelity. For instance, experimental results indicate that SFQ pulses can travel over 7~mm-long PTLs without re-amplification~\cite{talanov2022propagation}. Moreover, the footprint and kinetic inductance of PTLs can be substantially modified by their topology, material composition, and fabrication process. By harnessing these characteristics in conjunction with our high-speed SFQ control logic circuitry, we anticipate achieving data density improvements of two orders of magnitude compared to the current state of the art. This can be accomplished while retaining the same number of metal (memory) layers as existing fabrication processes. The method of vertical growth---common in technologies like CMOS V-NANDs---is also anticipated to increase data density gains by an additional order of magnitude.

Looking forward, although there are no fundamental limitations to the development of scalable superconducting delay-line memories, a direction that invites further exploration is that of impedance-matching transformers, such as tapers~\cite{4051841}. Alternatively, in order to address any potential impedance-mismatch challenges, exploring methods to boost the transmission line capacitance and utilizing higher-permittivity dielectrics, like NbOx~\cite{zhao2004optical}, between the signal and ground planes could prove to be a promising approach. Lastly, there are several interesting questions pertaining to the microarchitecture-compiler relationship. \rebuttal{Specifically, optimizations in hierarchical design, instruction scheduling, and data placement can significantly reduce the memory readout time and improve its efficiency in cases involving both sequential access and content-based addressing}~\cite{4038826}.

\section*{Methods}
\subsection*{Memory Control Circuit Design and Analysis}

\textbf{Simulation setup:} The simulations were conducted using WRSPICE~\cite{Wrspice}, a version of SPICE designed for superconductor electronic designs. The \rebuttal{resistively and capacitively-shunted JJ} model employed in the simulations is based on MIT Lincoln Laboratory's SFQ5ee 100~$\mu A/\mu m^2$ fabrication process~\cite{7386652} \rebuttal{with an $I_C R_n$ of 1.65~mV}. To generate SFQ input pulses from an input DC current stimulation, DC-to-SFQ converters were utilized. \rebuttal{The subsequent analysis does not consider parasitic inductances associated with shunt resistors, as their impact on circuit performance has been determined to be negligible~\cite{maezawa}.}\\

\noindent \textbf{Circuit design:} For the design of the memory controller's key components, as illustrated in Figures~\ref{fig:schem1}-\ref{fig:schem3}, we began with publicly accessible schematics~\cite{80745, dro2rdesign}. To achieve a higher operational frequency and wider bias margins, the number of JJs on the critical path of the corresponding circuit was minimized. To accomplish this, we designed a DRO cell with 4 JJs that matched the critical currents of JJs in the neighboring cells, resulting in improved signal quality. Additionally, the original DRO2R design features output paths that are symmetrical. However, as depicted in Figure~\ref{fig:arch2}, only the Q0 output (\emph{data\_out0}) is connected to the delay line loop. To meet the timing demands of the loop, the internal path from \emph{data\_in} to \emph{data\_out0} was shortened. The serial inductors in the Merger design were also reduced from their initial values to improve critical path latency. Finally, we eliminated active Splitter cells~\cite{80745}, which are commonly used to provide fan-out for shared nodes but incur an overhead of 3 JJs each. Instead, a recently-proposed $I_C$ ranking technique that allows for passive fan-out was adopted~\cite{10075073}. The combined modifications resulted in a memory controller circuit that comprises 29~JJs and a logic path delay of $\sim$10~ps, which is determined by the sum of the DRO2R cell's setup and propagation delays, averaged over its bias margins.\\

\noindent \textbf{Analysis:} To evaluate performance, the timing resolution was set to 0.5 ps, which was interpolated from the internal step size used by WRSPICE, and delays were measured as peak-to-peak values. Specifically, the DRO and DRO2R propagation delays were measured as the clock-to-Q delay, and the Merger propagation delay was measured as the delay from either input to the output. To determine the upper and lower time bounds of each cell, detailed bias margin analyses were conducted. For bias margin measurements, we started at the nominal voltage and decremented by steps of 1\% of the nominal voltage over the bias range where the circuit operates, to locate the lower limit. We then incremented the nominal voltage to locate the upper limit. Static timing analysis using minimum and maximum intervals was performed to ensure that the timing at the limits of the bias margins meets the hold, setup, and propagation time requirements at different target frequencies.

\subsection*{Memory Density Analysis}
The memory density of the PTL delay line was calculated using the equation:
\begin{equation}
Density = \frac{f}{w \times v}
\end{equation}
where $f$ represents the operating frequency of the memory controller in Hz, $w$ is the PTL pitch, and $v$ is the travel speed of single flux quanta in m/s. \\

\noindent \textbf{Controller operational frequency and PTL pitch:}
The delay results shown in Figure~\ref{fig:bias_vs_delay} indicate a maximum controller operational frequency of 100~GHz. In the conducted memory density analysis, as summarized in Table~\ref{tab:tab1}, the value of $f$ is selected to span from this value down to 20~GHz to highlight the trade-off between density versus bias margins, as suggested by the findings presented in Figure~\ref{fig:system_bias}. Regarding the PTL pitch---the sum of the PTL linewidth and minimum spacing requirement---this is typically determined by the fabrication process. For example, the well-established SFQ5ee~\cite{7386652} process allows for the reliable fabrication of Nb and MoN microstrips and striplines with a 250~nm linewidth and 250~nm spacing, or a pitch of 500~nm. The more advanced SC2~\cite{Tolpygo_2021} process can reduce the pitch to 240 nm. Further miniaturization down to a pitch of 135~nm and smaller is possible using e-beam lithography~\cite{PhysRevApplied.11.044014}. \\

\noindent \textbf{Travel speed:} The velocity factor equation was used to determine the travel speed ($v$) of single flux quanta in PTLs constructed with different materials. The equation is given by:
\begin{equation}
v = \frac{1}{c\sqrt{L \times C}}
\end{equation}
Here, $c$ is the speed of light in m/s, $L$ is the inductance per unit length in H/m, and $C$ is the capacitance per unit length in F/m. The capacitance and inductance values used in the calculations were sourced from literature and are shown in Table~\ref{tab:tab1}. In cases where geometric inductance is significant, such as in Nb striplines, the inductance scales non-linearly and must be evaluated at each linewidth individually. Based on available experimental measurements~\cite{Tolpygo_2021}, a 250~nm-wide Nb stripline has 0.5~pH/$\mu$m, while a 120~nm-wide line has 0.65~pH/$\mu$m. For high-kinetic inductors like MoN and NbTiN striplines and MoN microstrips, the kinetic inductance is \rebuttal{directly proportional to the length of the inductor and inversely proportional to its width}. This can be calculated by multiplying the known inductance per square values of 8~pH/$\square$ for MoN striplines and microstrips~\cite{8307498} and 49~pH/$\square$ for NbTiN striplines~\cite{PhysRevLett.122.010504} by the unit length divided by the linewidth. Regarding their capacitances, we relied on recently-published fabrication results~\cite{Tolpygo_2021}. For instance, a stripline having a linewidth of 120~nm has a capacitance of 0.19~fF/$\mu$m, which is solely determined by the dielectric material and geometry. Similarly, a microstrip having a linewidth of 120~nm has a capacitance of 0.14~fF/$\mu$m. For the 40~nm NbN nanowires, experimental measurements yielded an inductance of 82~pH/$\square$ or 2,050~pH/$\mu$m for this width, and a capacitance of 0.044~fF/$\mu$m~\cite{PhysRevApplied.11.044014}.Lastly, inductance and capacitance values for 15~nm NbN nanowires were obtained from simulation results~\cite{PhysRevApplied.11.044014}. \rebuttal{Effects of coherent quantum phase slips (CQPS) have been observed in NbN films with comparable cross-sectional dimensions~\cite{cqps}. 
Nevertheless, these effects do not present an issue here, as the kinetic inductance of the films, even with the most extreme linewidths considered, is still one order of magnitude lower than what has been used to observe CQPS. Consequently, the phase-slip amplitude in our case will be $e^{-9}$ times lower than in systems in which CQPS has been shown to be a significant contributor to conductivity~\cite{cqps2}
, and for this reason, we neglect it in our analysis.}

\noindent \textbf{Layer stacking:} An approach to further enhance memory density is by vertically stacking multiple layers of PTLs. In this case, the memory density increases linearly with the number of PTL layers, denoted by $N$ in the equation that follows:
\begin{equation}
Density = \frac{f}{w \times v}\times N
\end{equation}

\noindent In this regard, a stackup of Nb striplines with nine planarized superconducting layers, stackable stud vias, self-shunted Nb/AlOx-Al/Nb Josephson junctions, and a single layer of MoN kinetic inductors has been successfully developed and currently serves as the de facto fabrication process~\cite{8314189, stud}. This stackup\rebuttal{, MIT Lincoln Laboratory's state-of-the-art SFQ5ee process,} allows for four memory layers based on the typical allocation of ground planes and signal planes. \rebuttal{Several advancements to this process are under development, including the addition of self-shunted junctions~\cite{Tolpygo_2020}, which use higher critical current density and can reduce the chip area of digital hardware by 50\%, and the introduction of two additional routing layers~\cite{10049082}}. Looking forward, foundries have recently announced plans to scale up to sixteen routing layers and one layer for NbTiN kinetic inductors~\cite{herr2023scaling}; to the best of our knowledge, however, there is no fundamental limitation that constrains vertical growth~\cite{80745, radparvar1995superconducting, villegirr2001nbn, baggetta2005new, villegier1999extraction}.

\section*{Data availability}
All data relevant to the study are available from the corresponding authors upon request.

\bibliography{ref}

\section*{Acknowledgements}
The authors would like to thank D. Scott Holmes for his helpful comments and valuable discussions.

\section*{Author contributions statement}
J.V. and G.T. designed the proposed memory system and performed the timing analysis. J.V. designed, simulated, and analyzed the corresponding circuits. J.V., A.W., and G.T. contributed to the memory density analysis and the writing of the manuscript. G.T. conceived the idea and, with T.S., oversaw the project. All authors reviewed the manuscript. 

\subsection*{Corresponding authors}
Correspondence to \href{mailto:jevolk@ucsb.edu}{Jennifer Volk} and \href{mailto:gtzimpra@umich.edu}{Georgios Tzimpragos}.

\section*{Ethics declarations}
\subsection*{Competing interests}
The authors declare no competing interests.

\begin{table}[ht]
\resizebox{\textwidth}{!}{
\begin{tabular}{|l|l|l|l|l|l|l|l|l|l|l|}
\hline
\multicolumn{1}{|c|}{\textbf{Device}}                                                           & \multicolumn{1}{c|}{\textbf{\begin{tabular}[c]{@{}c@{}}Linewidth \\ (nm)\end{tabular}}} & \multicolumn{1}{c|}{\textbf{\begin{tabular}[c]{@{}c@{}}Spacing\\  (nm)\end{tabular}}} & \multicolumn{1}{c|}{\textbf{\begin{tabular}[c]{@{}c@{}}Fabrication\\ Process\end{tabular}}} & \multicolumn{1}{c|}{\textbf{\begin{tabular}[c]{@{}c@{}}Memory\\ Layers\end{tabular}}} & \multicolumn{1}{c|}{\textbf{\begin{tabular}[c]{@{}c@{}}Process\\ Maturity\end{tabular}}} & \multicolumn{1}{c|}{\textbf{\begin{tabular}[c]{@{}c@{}}Capacitance\\ (fF/$\mu$m)\end{tabular}}} & \multicolumn{1}{c|}{\textbf{\begin{tabular}[c]{@{}c@{}}Inductance\\ (pH/$\mu$m)\end{tabular}}} & \multicolumn{1}{c|}{\textbf{\begin{tabular}[c]{@{}c@{}}Travel\\ Speed ($\times$c)\end{tabular}}} & \multicolumn{1}{c|}{\textbf{\begin{tabular}[c]{@{}c@{}}Frequency\\ (GHz)\end{tabular}}} & \multicolumn{1}{c|}{\textbf{\begin{tabular}[c]{@{}c@{}}Density\\ (Mbit/cm$^2$)\end{tabular}}} \\ \hline
\multirow{8}{*}{\begin{tabular}[c]{@{}l@{}}Nb\\ Stripline\end{tabular}}                         & \multirow{4}{*}{250}                                                                    & \multirow{4}{*}{250}                                                                  & \multirow{4}{*}{SFQ5ee}                                                                     & \multirow{4}{*}{4}                                                                            & \multirow{4}{*}{Mature}                                                                  & \multirow{4}{*}{0.25}                                                                           & \multirow{4}{*}{0.50}                                                                          & \multirow{4}{*}{0.298}                                                                           & 20                                                                                      & 0.2                                                                                            \\ \cline{10-11} 
                                                                                                &                                                                                         &                                                                                       &                                                                                             &                                                                                               &                                                                                          &                                                                                                 &                                                                                                &                                                                                                  & 50                                                                                      & 0.4                                                                                            \\ \cline{10-11} 
                                                                                                &                                                                                         &                                                                                       &                                                                                             &                                                                                               &                                                                                          &                                                                                                 &                                                                                                &                                                                                                  & 75                                                                                      & 0.7                                                                                            \\ \cline{10-11} 
                                                                                                &                                                                                         &                                                                                       &                                                                                             &                                                                                               &                                                                                          &                                                                                                 &                                                                                                &                                                                                                  & 100                                                                                     & 0.9                                                                                            \\ \cline{2-11} 
                                                                                                & \multirow{4}{*}{120}                                                                    & \multirow{4}{*}{120}                                                                  & \multirow{4}{*}{SC2}                                                                        & \multirow{4}{*}{4}                                                                            & \multirow{4}{*}{Aggressive}                                                               & \multirow{4}{*}{0.19}                                                                           & \multirow{4}{*}{0.65}                                                                          & \multirow{4}{*}{0.296}                                                                           & 20                                                                                      & 0.4                                                                                            \\ \cline{10-11} 
                                                                                                &                                                                                         &                                                                                       &                                                                                             &                                                                                               &                                                                                          &                                                                                                 &                                                                                                &                                                                                                  & 50                                                                                      & 0.9                                                                                            \\ \cline{10-11} 
                                                                                                &                                                                                         &                                                                                       &                                                                                             &                                                                                               &                                                                                          &                                                                                                 &                                                                                                &                                                                                                  & 75                                                                                      & 1.4                                                                                            \\ \cline{10-11} 
                                                                                                &                                                                                         &                                                                                       &                                                                                             &                                                                                               &                                                                                          &                                                                                                 &                                                                                                &                                                                                                  & 100                                                                                     & 1.9                                                                                            \\ \hline
\multirow{8}{*}{\begin{tabular}[c]{@{}l@{}}MoN \\ Kinetic\\ Inductor\\ Microstrip\end{tabular}} & \multirow{4}{*}{250}                                                                    & \multirow{4}{*}{250}                                                                  & \multirow{4}{*}{SFQ5ee}                                                                     & \multirow{4}{*}{1}                                                                            & \multirow{4}{*}{Mature}                                                                  & \multirow{4}{*}{0.16}                                                                           & \multirow{4}{*}{32}                                                                            & \multirow{4}{*}{0.047}                                                                           & 20                                                                                      & 0.3                                                                                            \\ \cline{10-11} 
                                                                                                &                                                                                         &                                                                                       &                                                                                             &                                                                                               &                                                                                          &                                                                                                 &                                                                                                &                                                                                                  & 50                                                                                      & 0.7                                                                                            \\ \cline{10-11} 
                                                                                                &                                                                                         &                                                                                       &                                                                                             &                                                                                               &                                                                                          &                                                                                                 &                                                                                                &                                                                                                  & 75                                                                                      & 1.1                                                                                            \\ \cline{10-11} 
                                                                                                &                                                                                         &                                                                                       &                                                                                             &                                                                                               &                                                                                          &                                                                                                 &                                                                                                &                                                                                                  & 100                                                                                     & 1.4                                                                                            \\ \cline{2-11} 
                                                                                                & \multirow{4}{*}{120}                                                                    & \multirow{4}{*}{120}                                                                  & \multirow{4}{*}{SC2}                                                                        & \multirow{4}{*}{1}                                                                            & \multirow{4}{*}{Aggressive}                                                               & \multirow{4}{*}{0.14}                                                                           & \multirow{4}{*}{66.70}                                                                         & \multirow{4}{*}{0.034}                                                                           & 20                                                                                      & 0.8                                                                                            \\ \cline{10-11} 
                                                                                                &                                                                                         &                                                                                       &                                                                                             &                                                                                               &                                                                                          &                                                                                                 &                                                                                                &                                                                                                  & 50                                                                                      & 2.0                                                                                            \\ \cline{10-11} 
                                                                                                &                                                                                         &                                                                                       &                                                                                             &                                                                                               &                                                                                          &                                                                                                 &                                                                                                &                                                                                                  & 75                                                                                      & 3.0                                                                                            \\ \cline{10-11} 
                                                                                                &                                                                                         &                                                                                       &                                                                                             &                                                                                               &                                                                                          &                                                                                                 &                                                                                                &                                                                                                  & 100                                                                                     & 4.0                                                                                            \\ \hline
\multirow{4}{*}{\begin{tabular}[c]{@{}l@{}}MoN\\ Kinetic\\ Inductor\\ Stripline\end{tabular}}   & \multirow{4}{*}{120}                                                                    & \multirow{4}{*}{120}                                                                  & \multirow{4}{*}{SC2}                                                                        & \multirow{4}{*}{4}                                                                            & \multirow{4}{*}{Aggressive}                                                               & \multirow{4}{*}{0.19}                                                                           & \multirow{4}{*}{66.70}                                                                         & \multirow{4}{*}{0.029}                                                                           & 20                                                                                      & 3.2                                                                                            \\ \cline{10-11} 
                                                                                                &                                                                                         &                                                                                       &                                                                                             &                                                                                               &                                                                                          &                                                                                                 &                                                                                                &                                                                                                  & 50                                                                                      & 8.1                                                                                            \\ \cline{10-11} 
                                                                                                &                                                                                         &                                                                                       &                                                                                             &                                                                                               &                                                                                          &                                                                                                 &                                                                                                &                                                                                                  & 75                                                                                      & 12.1                                                                                           \\ \cline{10-11} 
                                                                                                &                                                                                         &                                                                                       &                                                                                             &                                                                                               &                                                                                          &                                                                                                 &                                                                                                &                                                                                                  & 100                                                                                     & 19.0                                                                                           \\ \hline
                                                                                                \multirow{4}{*}{\begin{tabular}[c]{@{}l@{}}NbTiN\\ Kinetic\\ Inductor\\ Stripline\end{tabular}} & \multirow{4}{*}{100}                                                                    & \multirow{4}{*}{120}                                                                  & \multirow{4}{*}{\begin{tabular}[c]{@{}l@{}}Not\\ Established\end{tabular}}                  & \multirow{4}{*}{4}                                                                            & \multirow{4}{*}{Academic}                                                                & \multirow{4}{*}{0.17}                                                                           & \multirow{4}{*}{490.5}                                                                        & \multirow{4}{*}{0.011}                                                                           & 20                                                                                      & 10.7                                                                                           \\ \cline{10-11} 
                                                                                                &                                                                                         &                                                                                       &                                                                                             &                                                                                               &                                                                                          &                                                                                                 &                                                                                                &                                                                                                  & 50                                                                                      & 26.6                                                                                           \\ \cline{10-11} 
                                                                                                &                                                                                         &                                                                                       &                                                                                             &                                                                                               &                                                                                          &                                                                                                 &                                                                                                &                                                                                                  & 75                                                                                      & 40.0                                                                                           \\ \cline{10-11} 
                                                                                                &                                                                                         &                                                                                       &                                                                                             &                                                                                               &                                                                                          &                                                                                                 &                                                                                                &                                                                                                  & 100                                                                                     & 53.3                                                                                           \\ \hline
\multirow{12}{*}{\begin{tabular}[c]{@{}l@{}}NbN\\ Kinetic\\ Inductor\\ Nanowire\end{tabular}}   & \multirow{4}{*}{40}                                                                     & \multirow{4}{*}{120}                                                                  & \multirow{4}{*}{\begin{tabular}[c]{@{}l@{}}Not\\ Established\end{tabular}}                  & \multirow{4}{*}{4}                                                                            & \multirow{4}{*}{Academic}                                                                & \multirow{4}{*}{0.04}                                                                           & \multirow{4}{*}{2,050}                                                                         & \multirow{4}{*}{0.011}                                                                           & 20                                                                                      & 15.1                                                                                           \\ \cline{10-11} 
                                                                                                &                                                                                         &                                                                                       &                                                                                             &                                                                                               &                                                                                          &                                                                                                 &                                                                                                &                                                                                                  & 50                                                                                      & 37.7                                                                                           \\ \cline{10-11} 
                                                                                                &                                                                                         &                                                                                       &                                                                                             &                                                                                               &                                                                                          &                                                                                                 &                                                                                                &                                                                                                  & 75                                                                                      & 56.6                                                                                           \\ \cline{10-11} 
                                                                                                &                                                                                         &                                                                                       &                                                                                             &                                                                                               &                                                                                          &                                                                                                 &                                                                                                &                                                                                                  & 100                                                                                     & 75.4                                                                                           \\ \cline{2-11} 
                                                                                                & \multirow{4}{*}{15}                                                                     & \multirow{4}{*}{120}                                                                  & \multirow{4}{*}{\begin{tabular}[c]{@{}l@{}}Not\\ Established\end{tabular}}                  & \multirow{4}{*}{4}                                                                            & \multirow{4}{*}{Academic}                                                                & \multirow{4}{*}{0.04}                                                                           & \multirow{4}{*}{5,467}                                                                         & \multirow{4}{*}{0.007}                                                                           & 20                                                                                      & 28.1                                                                                           \\ \cline{10-11} 
                                                                                                &                                                                                         &                                                                                       &                                                                                             &                                                                                               &                                                                                          &                                                                                                 &                                                                                                &                                                                                                  & 50                                                                                      & 70.1                                                                                           \\ \cline{10-11} 
                                                                                                &                                                                                         &                                                                                       &                                                                                             &                                                                                               &                                                                                          &                                                                                                 &                                                                                                &                                                                                                  & 75                                                                                      & 105.2                                                                                          \\ \cline{10-11} 
                                                                                                &                                                                                         &                                                                                       &                                                                                             &                                                                                               &                                                                                          &                                                                                                 &                                                                                                &                                                                                                  & 100                                                                                     & 140.3                                                                                          \\ \cline{2-11} 
                                                                                                & \multirow{4}{*}{15}                                                                     & \multirow{4}{*}{120}                                                                  & \multirow{4}{*}{\begin{tabular}[c]{@{}l@{}}Not \\ Established\end{tabular}}                 & \multirow{4}{*}{100}                                                                          & \multirow{4}{*}{Academic}                                                                & \multirow{4}{*}{0.04}                                                                           & \multirow{4}{*}{5,467}                                                                         & \multirow{4}{*}{0.007}                                                                           & 20                                                                                      & 701.4                                                                                          \\ \cline{10-11} 
                                                                                                &                                                                                         &                                                                                       &                                                                                             &                                                                                               &                                                                                          &                                                                                                 &                                                                                                &                                                                                                  & 50                                                                                      & 1,753                                                                                          \\ \cline{10-11} 
                                                                                                &                                                                                         &                                                                                       &                                                                                             &                                                                                               &                                                                                          &                                                                                                 &                                                                                                &                                                                                                  & 75                                                                                      & 2,630                                                                                          \\ \cline{10-11} 
                                                                                                &                                                                                         &                                                                                       &                                                                                             &                                                                                               &                                                                                          &                                                                                                 &                                                                                                &                                                                                                  & 100                                                                                     & 3,507                                                                                          \\ \hline
\end{tabular}
}
\caption{Memory density estimates for a variety of mature, aggressive, and academic fabrication processes. The MIT Lincoln Laboratory SFQ5ee 100~$\mu$A/$\mu$m$^2$ process has served as the de facto standard for fabrication since its introduction in 2016, and thus is considered mature~\cite{7386652}. The MIT Lincoln Laboratory SC2~\cite{Tolpygo_2021} is considered aggressive, as it has yielded successful experimental circuits, but pushes the bounds on common design rules. The term academic refers to processes that have been purpose-built and evaluated either experimentally or theoretically.}

\label{tab:tab1}
\end{table}

\end{document}